\documentclass{aa}
\usepackage[dvips]{graphicx}

\begin{document}

\title{IP Pegasi: Investigation of the accretion disk structure
\thanks{Based on observations made at the Special Astrophysical Observatory,
Nizhnij Arkhyz, Russia, and at the German-Spanish Astronomical
Center, Calar Alto, Spain.}}
\subtitle{Searching evidences for spiral shocks in the quiescent accretion disk}

\author{V.V. Neustroev \inst{1}, N.V. Borisov \inst{2}, H. Barwig \inst{3},
A. Bobinger \inst{3}, K.H. Mantel \inst{3}, D. \v{S}imi\'{c} \inst{3}, S. Wolf \inst{3}}

\offprints{V.V. Neustroev (\email{benj@uni.udm.ru})}

\institute{
           Department of Astronomy and Mechanics, Udmurtia State University,
           Universitetskaya str., 1, Izhevsk, 426034, Russia
 \and
           Special Astrophysical Observatory, Nizhnij Arkhyz,
           Karachaevo-Cherkesia, 357147, Russia
 \and
           Universit\"{a}ts­-Sternwarte, Scheinerstr. 1, D-81679 M\"{u}nchen, Germany
}

\date{Received 7 May 2002 / Accepted 8 July 2002}

\titlerunning{Searching evidences for spiral shocks in the quiescent accretion disk of IP Peg}
\authorrunning{V.V. Neustroev et al.}

\abstract{
We present the results of spectral investigations of the cataclysmic variable
IP Peg in quiescence.
Optical spectra obtained on the 6-m telescope at the Special Astrophysical Observatory
(Russia), and on the 3.5-m telescope at the German-Spanish Astronomical
Center (Calar Alto, Spain), have been analysed by means of Doppler tomography and
Phase Modelling Technique. From this analysis we conclude that
the quiescent accretion disk of IP Peg has a complex structure.
There are also explicit indications of spiral shocks.
The Doppler maps and the variations of the peak separation of the emission lines
confirm this interpretation.
We have detected that all the emission lines show a
rather considerable asymmetry of their wings varying with time. The wing asymmetry
shows quasi-periodic modulations with a period much shorter than the orbital one.
This indicates the presence of an emission source in the binary rotating asynchronously
with the binary system.
We also have found that the brightness of the bright spot changes considerably
during one orbital period. The spot becomes brightest at an inferior conjunction,
whereas it is almost invisible when it is located on the distant
half of the accretion disk.
Probably, this phenomenon is due to an anisotropic radiation of the bright
spot and an eclipse of the bright spot by the outer edge of the accretion disk.

\keywords{Accretion, accretion disk -- line: profiles -- line: formation --
    novae, cataclysmic variables -- stars: individual: IP Peg}
}

\maketitle

\section{Introduction}

IP Pegasi is a well-known deeply eclipsing dwarf nova with an orbital period of 3\fh79
(Lipovetskij \& Stephanyan \cite{Lipov:Steph}; Wolf et al. \cite{Wolf93}) and an
outburst cycle of about 2.5 months. During high states, which last about 12 days,
the system changes from V magnitude 14th during quiescence (out of eclipse) to 12
during outburst. Many times IP Peg was subject to detailed spectral investigations
(Marsh \& Horne \cite{marsh:horne90};
Marsh \cite{marsh88}; Martin et al. \cite{Martin}; Hessman \cite{Hessman};
Harlaftis et al. \cite{Harlaftis94}; Wolf et al. \cite{Wolf98})
and photometric researches (Wolf et al. \cite{Wolf93}; Szkody \& Mateo \cite{Szkody};
Wood \& Crawford \cite{Wood86};
Wood et al. \cite{Wood89}; Bobinger et al. \cite{Bobinger97}).
In the last years IP Peg deserved increasing interest as
spiral shocks were detected in its accretion disk during outburst
(Steeghs et al. \cite{Steeghs97}; Harlaftis et al. \cite{Harlaftis99}).

The question on the existence of spiral shocks in the accretion disks has
a long standing history. It is closely related to
angular momentum transport mechanisms in accretion disks.
At present there are two approaches to this problem: According to the
first one, angular momentum is transported due to the presence of turbulent
or magnetic viscosity in the disk (Shakura \& Sunyaev \cite{Shak:Sun}).
On the other hand, hydrodynamical
numerical calculations have shown that tidal forces of the secondary
induce spiral shock waves in the accretion disk, which may provide
an efficient transfer mechanism (Sawada et al.~\cite{Japan86};
Matsuda et al.~\cite{Japan90}).
Self-similar solutions having spiral shocks, was constructed in a
semi-analytic manner in two dimensions by Spruit (\cite{Spruit},
see also Chakrabarti \cite{Chak}).
It should be noted that these two mechanisms are mutually exclusive,
as shock waves in the presence of viscosity will be smeared out
(Bunk et al. \cite{Bunk}; Chakrabarti \cite{Chak}).
Although Steeghs et al. (1997) found indications for spiral shocks in the hot accretion
disks during an outburst, the problem on the spiral structure of the quiescent
accretion disks still remains unsolved.

Most theorists do not support the hypothesis of the presence of
spiral shocks in the accretion disks in quiescence
due to the following arguments:

\noindent Since the disk expands during outburst (Smak \cite{Smak84};
Wolf et al. \cite{Wolf93}), the outer parts of the expanded disk are
subject to higher gravitational attraction of the secondary,
leading to the formation of spiral arms. The tidal force is a very
steep function of the distance to the secondary star, hence the spiral
shocks rapidly become weak at smaller disk sizes.\footnote{However,
it should be noted that, while the tidal force decreases
with distance, the spiral shocks can amplify within the disk as
it travels inwards (Spruit \cite{Spruit}).}
Furthermore, the opening angle of the spirals is directly related to the
temperature of the disk as it roughly propagates at sound speed.
This effect was the principal reason to believe that spirals may
not be present in accretion disks of cataclysmic variables, which
are not hot enough. This is in principal the case in quiescent disks
(Boffin \cite{Boffin01}). Steeghs \& Stehle found from their calculations
(\cite{Stehle99}), using the grid of the 2D hydrodynamic accretion disk,
that the spiral shocks in quiescent accretion disks are so
tightly wound that they leave few fingerprints in the emission lines.

At the same time, the authors of the most comprehensive 3D simulations are
in principle inclined to accept the presence of
spiral shocks in the relatively cold disks. So, Makita et al. (\cite{Japan00})
observed in their 3D calculations that spiral shocks are formed in an accretion
disk for all specific heat ratios. In addition, they claim that in cool disks,
when comparing 3D calculations with 2D ones, spiral arms are less tightly
wound in the outer region of the disk.
In connection with this, it is important to add that in 3D disks,
there is a vertical resonance that can lie within the disk.
This resonance causes a local thickening of the disk and generates
waves that propagate radially. The first wavelength of this wave
is longer than that of the usual 2D waves by a factor of about
$(R/H)^{1/3}$ (Lubow \cite{Lubow}). The wavelength does increase with
the level of non-linearity (Yuan \& Cassen \cite{Yuan:Cassen}).
The non-linear effects can modify significantly the appearance of the spiral
pattern: a longer wavelength produces a less tightly wound spiral pattern.
Obviously the observations matches more closely a 3D disk, and the
2D approximation can poorly explain the observations.
[Indeed, the two-dimensional disk simulations demonstrate that the spiral pattern
resembles the observations only for unreally high temperatures
(see, for example, Godon et al. \cite{Godon}).]
And at last, the calculations of Bisikalo et al.
(\cite{Bisikalo98a,Bisikalo98b}, see also Kuznetsov et al. \cite{Kuznetsov})
only yield the one-armed spiral shock in the quiescent accretion disk.
In the place where the second arm should form, the stream from $L_{1}$ dominates and
presumably prevents the formation of the second arm of the tidally induced
spiral shock.

Thus, as stated above, the existence of the spiral shocks in the quiescent accretion
disk has not been established convincingly. Their observational detection would
be very important because spiral arms are indeed very efficient in transporting angular
momentum into the outer part of the disk. If existing at all, spiral shocks
would be much more difficult to detect than the strong shocks in the hot
accretion disk during outburst. Nevertheless, first steps towards an
observational confirmation have been started.
So, studying the structure of the accretion disk of U Gem in quiescence we have
detected sinusoidal variations ($\sin 2\varphi$) of the double peak separations
of the hydrogen emission lines and of their shape
(Borisov \& Neustroev \cite{bor:neus2}).
These variations are the signature of an m=2 mode in the disk.
This mode can be excited by the tidal forcing.
We showed that such orbital behavior
of the emission lines can be reproduced by means of a simple model of the accretion disk
with two symmetrically located spiral shocks (Neustroev \& Borisov \cite{Neus:Bor}).
Recently we have found, by means of Doppler tomography, new confirmation
for spiral shocks in the quiescent accretion disk of U~Gem (Neustroev et al.
\cite{Neus2002}). Further investigations in this area are strongly recommended.

The main goal of the present paper was the investigation of the accretion disk structure
of IP Peg in order to locate the line emitting sites during quiescence.
Special attention was payed to the search for evidences for a spiral
structure of the accretion disk.
Section~\ref{observations} of this paper describes the observations and data
reduction procedures, while Sect.~\ref{average} describes an average spectrum and
emission lines variations.
Section~\ref{structure} presents the analysis of
the optical spectra by means of Doppler tomography and Phase Modelling Technique.
The results are discussed in Sect.~\ref{Discussion} and
summarized in Sect.~\ref{Conclusions}.

\begin{figure}
\resizebox{\hsize}{!}{
\includegraphics{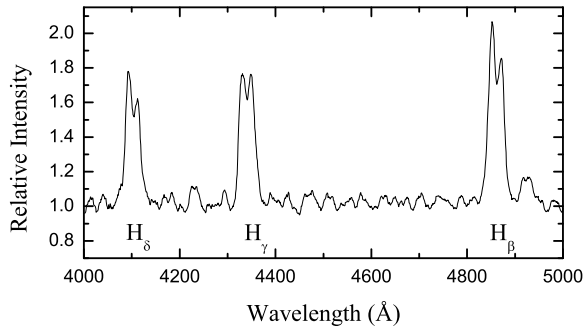}
}
\caption{The mean normalized spectra of IP Peg, based on first dataset.}
\label{ave_spec}
\end{figure}

\begin{figure*}
\flushright
\resizebox{13.5cm}{!}{
\includegraphics{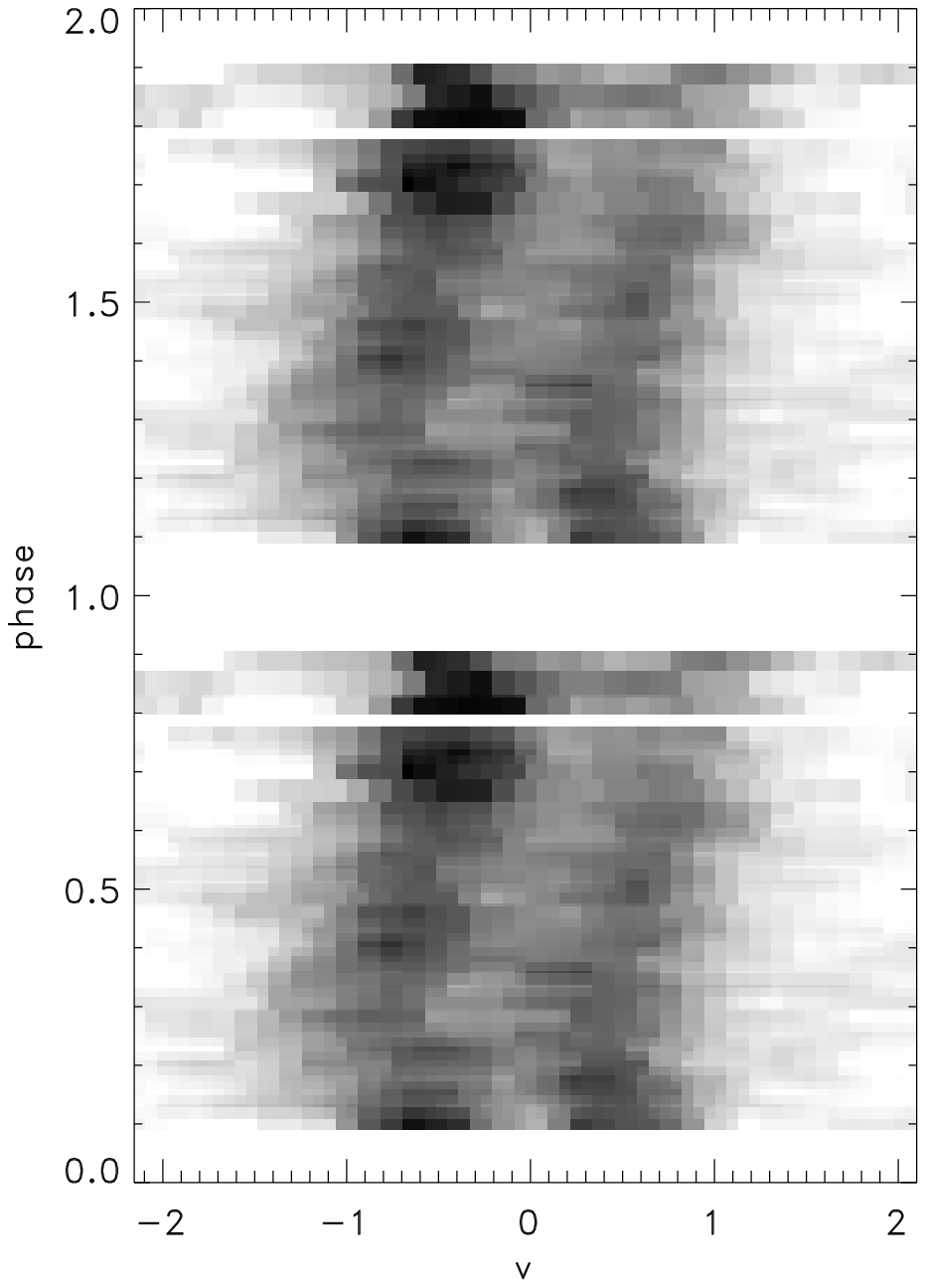}
\includegraphics{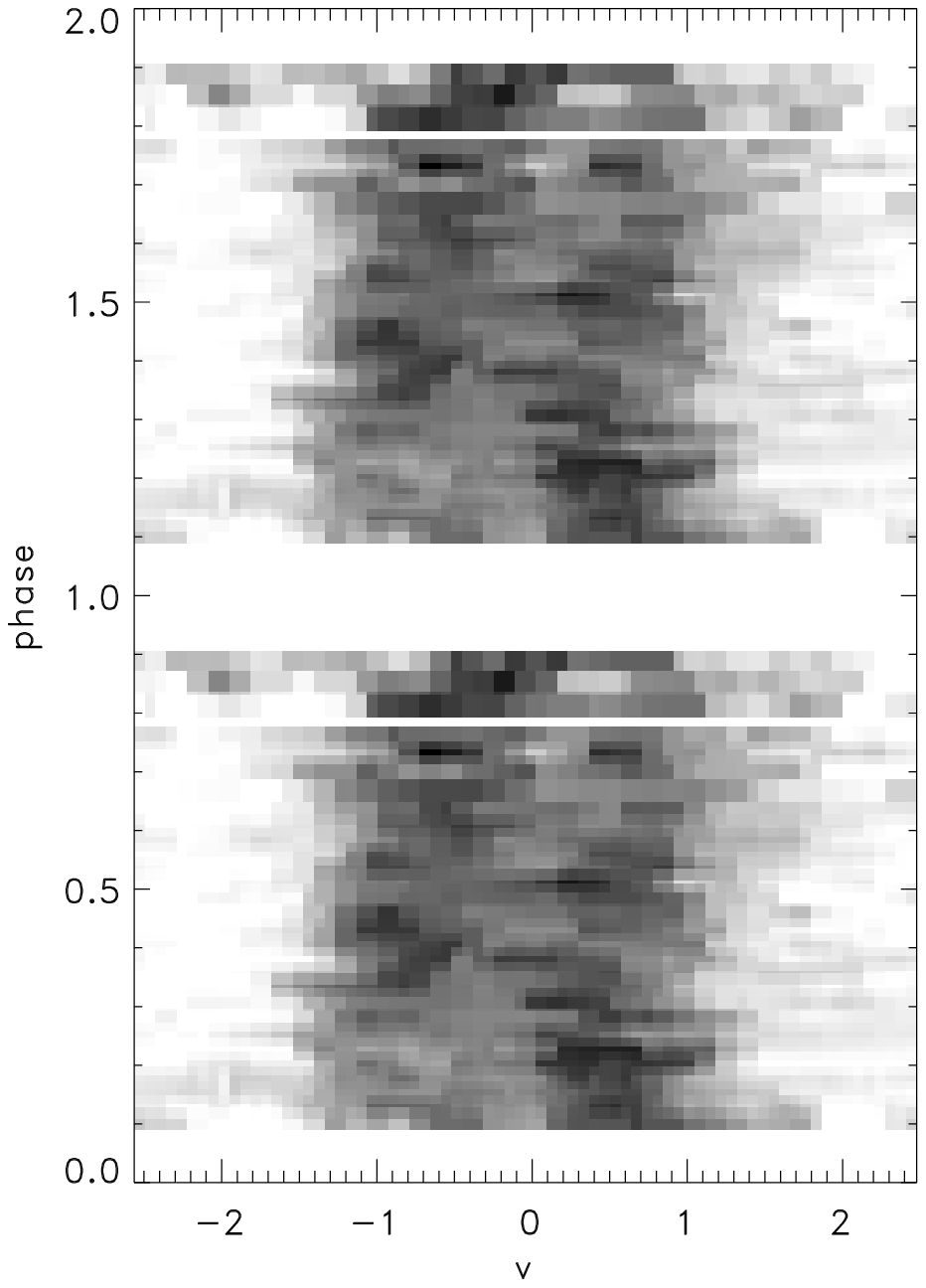}
\includegraphics{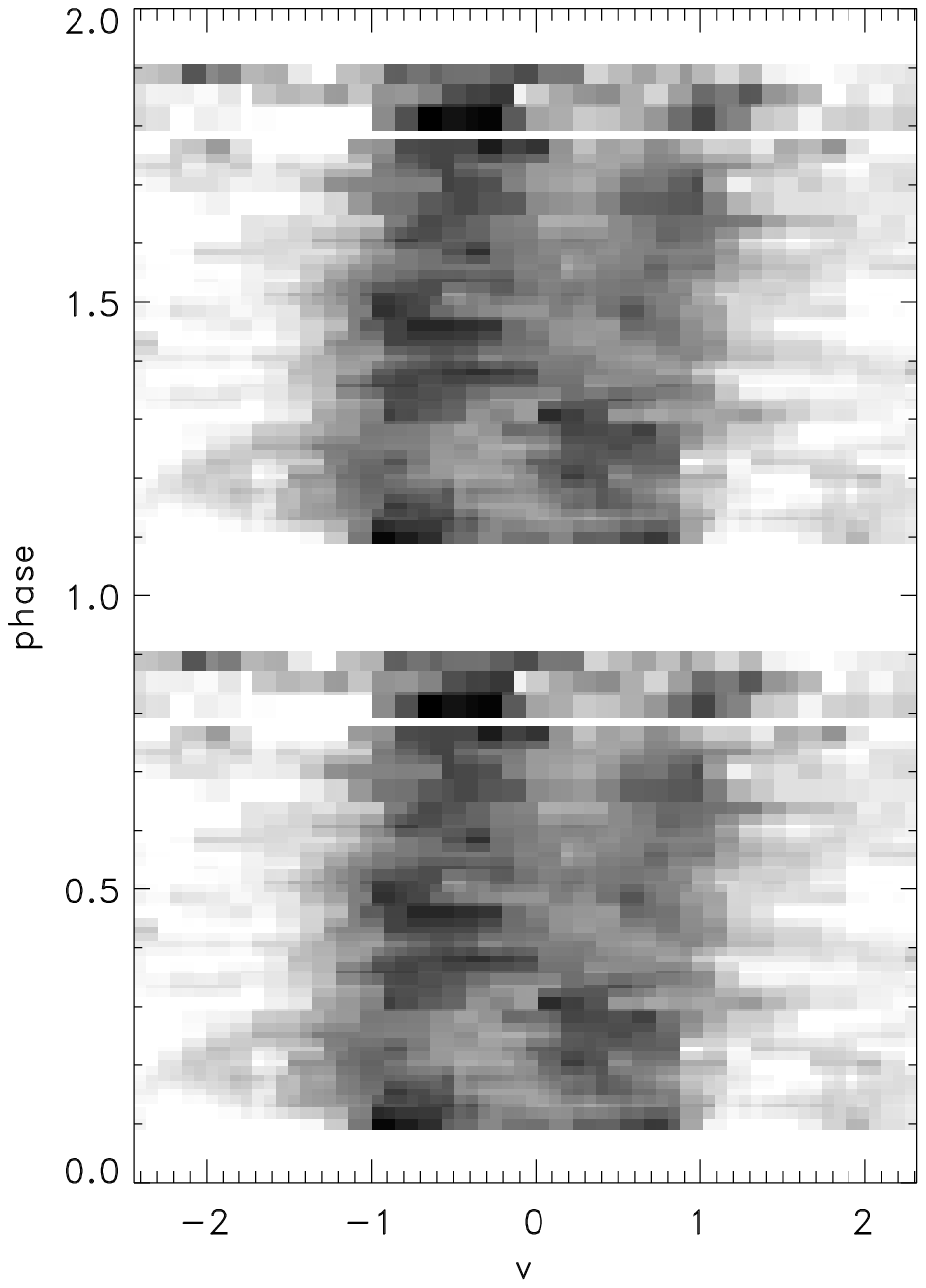}
}
\hfill
\\
\vspace*{-0.5cm}
\flushleft
\resizebox{13.5cm}{!}{
\includegraphics{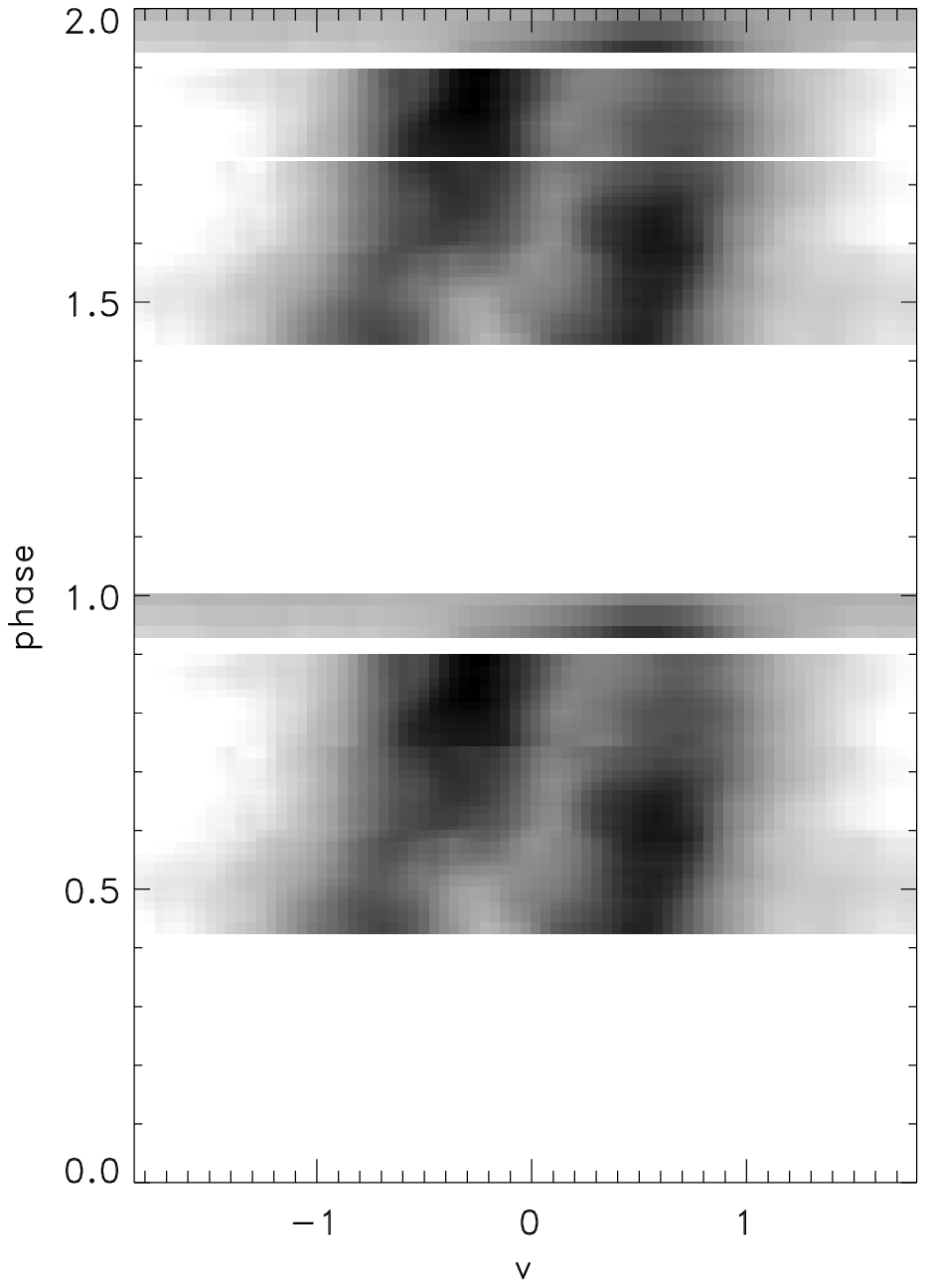}
\includegraphics{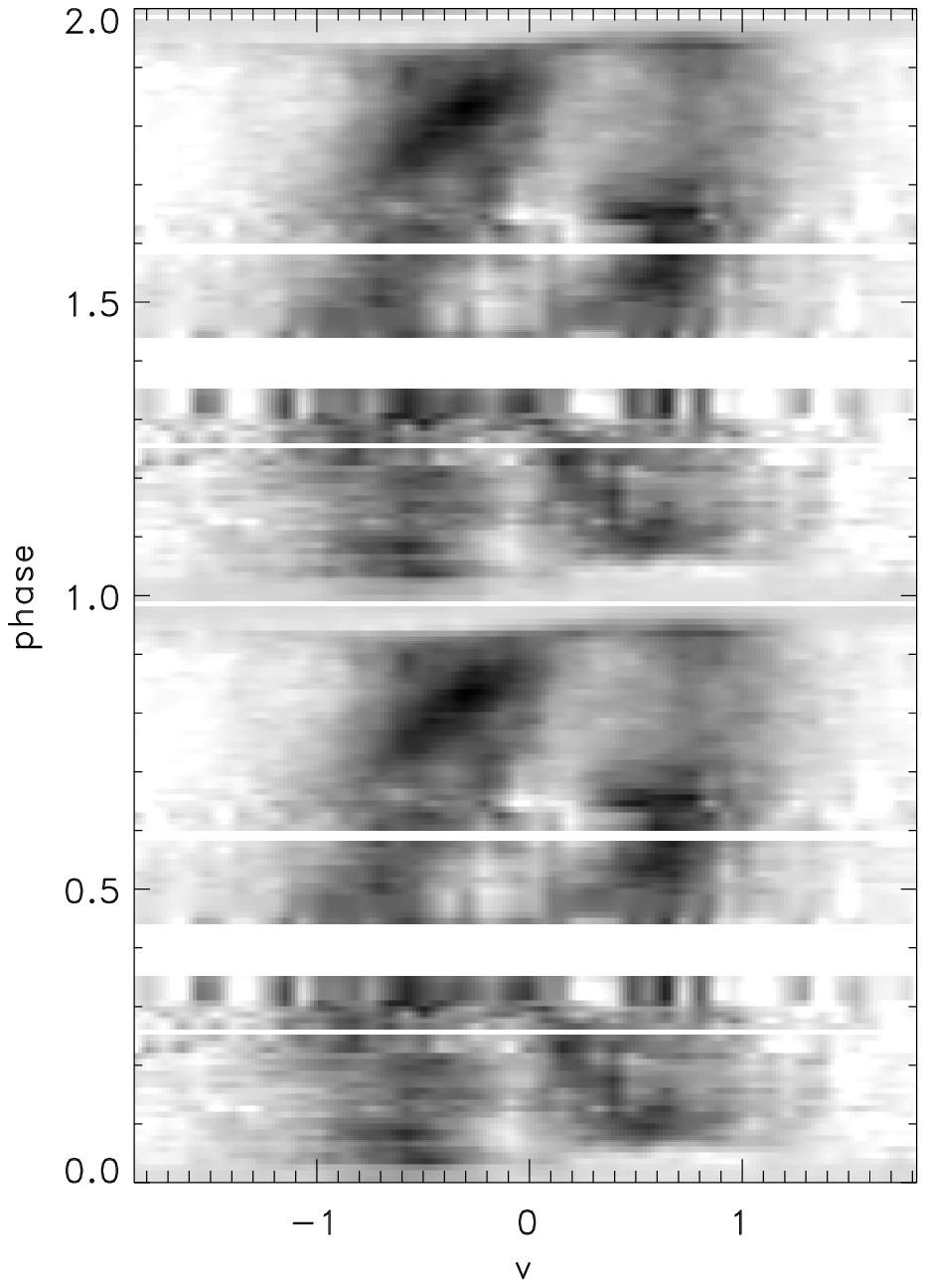}
\includegraphics{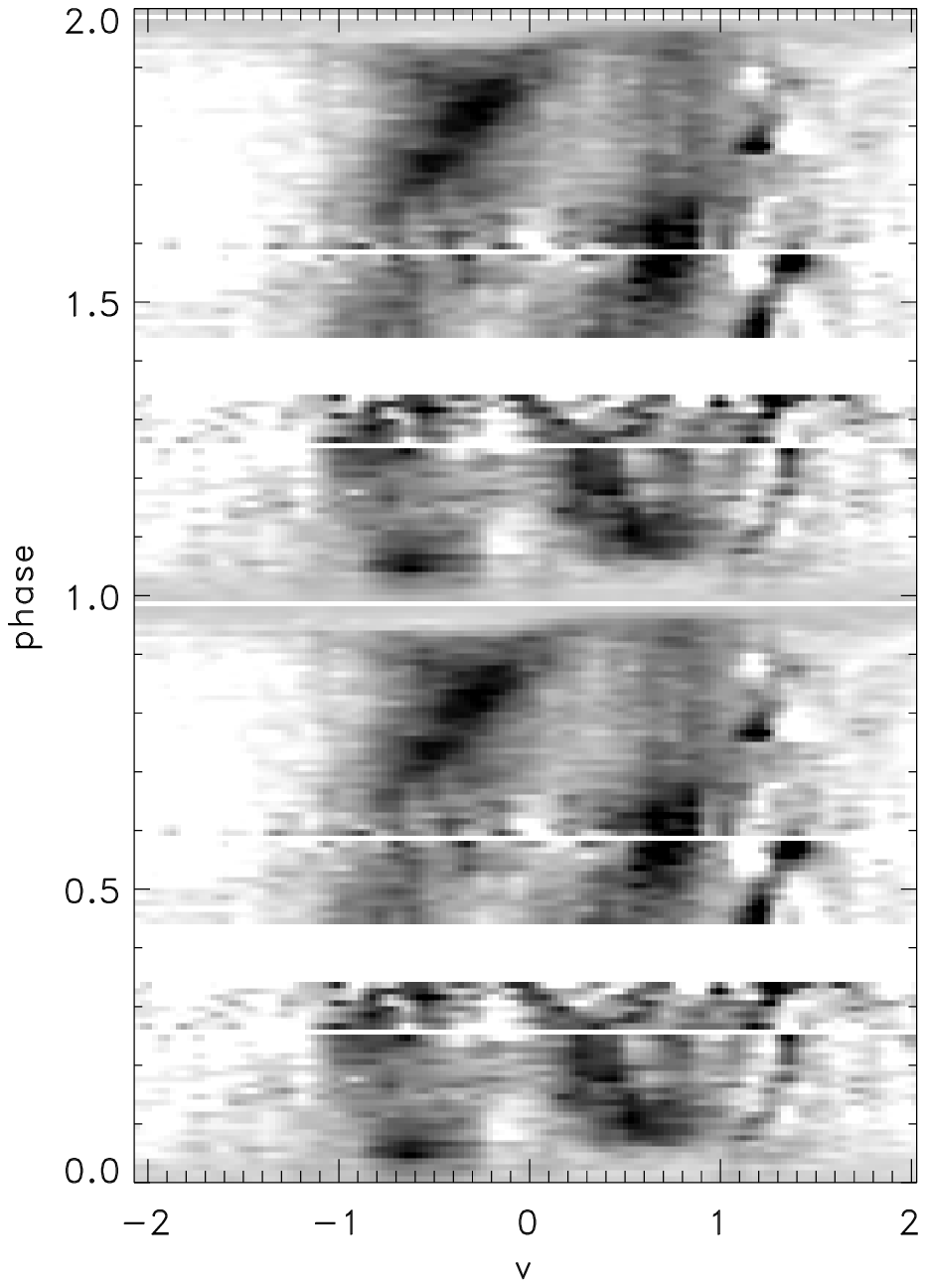}
}
\caption{The trailed spectra of the major emission lines.
\textit{Top row:} From left to right, the H$\beta$, H$\gamma$ and H$\delta$ lines from
the first dataset are presented. \textit{Bottom row:} From left to right, the H$\alpha$,
H$\beta$ and H$\gamma$ lines from the second dataset are presented. Note,
that we could only obtain 50\% phase coverage of H$\alpha$ due to technical
difficulties.}
\label{trailed}
\end{figure*}

\section{Observations}
\label{observations}

The spectra presented here were obtained during two observing runs.
The first spectroscopic observations of IP Peg were performed on
July 13, 1993 using a 1000-channel television scanner mounted on the SP-124
spectrograph at the Nasmyth-1 focus of the 6-m telescope of the Special
Astrophysical Observatory. A total of 40 spectra were taken in the wavelength range
3600-5500 {\AA} with a dispersion of 1.9 {\AA} channel$^{-1}$.
The weather conditions were not very favorable, with a seeing of around
3 arcseconds. Individual exposure times were 300 s, and the intervals between
them did not exceed 30 s. The total duration of the observations was 3.8 hours.
Note that we observed IP Peg in its quiescent state. We reduced these spectroscopic
data by using the procedure described by Knyazev (\cite{Knyazev}).

A second dataset was obtained during an observing campaign in August 06-09
1994 at the 3.5-m telescope at the German-Spanish Astronomical Center,
Calar Alto, Spain. From the trailed spectra, which were phase-folded into
100 phase bins and which have a spectral resolution of 1.3 \AA, we extracted
the H$\alpha$, H$\beta$ and H$\gamma$ lines. Due to technical difficulties
we could only obtain 50\% phase coverage of H$\alpha$. Data acquisition and
reduction methods used are described in detail by Wolf et al. (\cite{Wolf98}).

\section{Average spectrum and emission lines variations}
\label{average}

The mean normalized spectrum of IP Peg, based on first dataset, is shown
in Fig.~\ref{ave_spec}. The mean spectrum for second dataset can be found
in Wolf et al. (\cite{Wolf98}, Fig.~2). Before
averaging the single spectra were shifted according to the $K_{1}\sin(\phi)$,
with $K_{1}$ = 168 km s$^{-1}$ (Wolf et al. \cite{Wolf98}).
For the calculation of orbital phases the ephemeris of Wolf et al.
(\cite{Wolf93}) was used:

$T_{0}(HJD) = 2445615.4156 + 0.15820616 \cdot E$,

\noindent where $T_{0}$ is the moment of mid-eclipse.

Both spectra show strong, broad hydrogen and helium emission lines.
The trailed spectra of the major hydrogen lines are shown in Fig.~\ref{trailed}
(note that the data are displayed twice for clarity).
Their double-peaked profiles are thought to be associated with the accretion
disk. They are superposed by asymmetric structures produced by
anisotropically radiating emission regions of the binary system.
The line profiles
demonstrate a puzzling behaviour the blue-shifted peak being stronger
than the red-shifted peak when averaged over the orbit.

\begin{figure}
\resizebox{\hsize}{!}{
\includegraphics{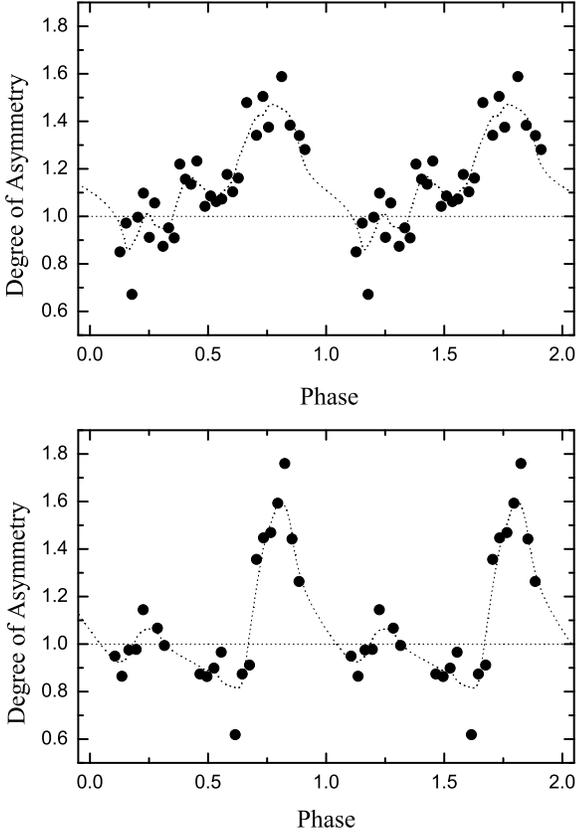}
}
\caption{The degree of asymmetry of the emission line H$\beta $ folded with
the orbital period (top panel: first dataset, bottom panel: second dataset).
The degree of asymmetry is the ratio between the
areas of violet and red peaks of the emission line.
Filled circles show individual values, open circles with dotted line show data
which were obtained by averaging adjacent data points.
The data are plotted twice for continuity.}
\label{swave}
\end{figure}

\begin{figure}
\resizebox{8cm}{!}{
\includegraphics{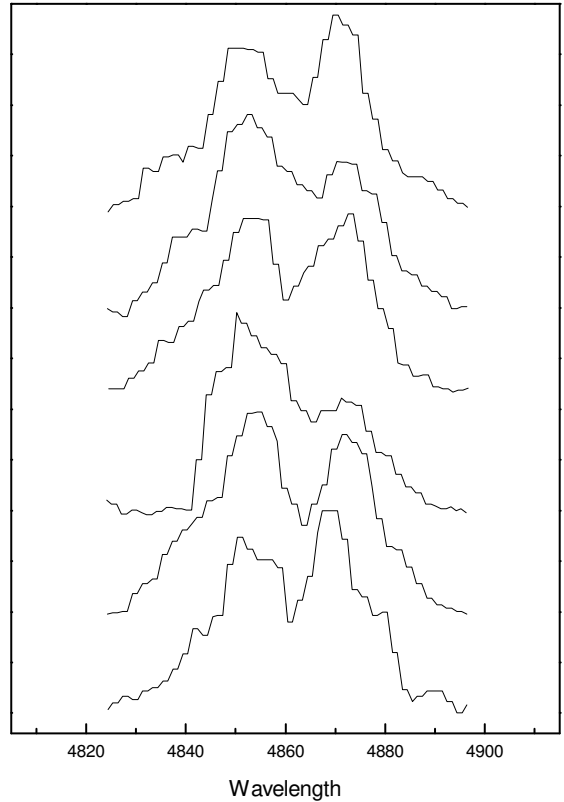}
}
\vspace{0.9 cm}
\caption{
Some profiles of the H$\beta$ line which show the rather considerable asymmetry of
its wings. It can be seen that from time to time on blue and red wings a hump appears.}
\label{wing_as}
\end{figure}

We have measured the degree of asymmetry of each line profile, and have
plotted this parameter against the orbital phase (in the following this plot
is named: the S-wave graph). We have calculated the degree of asymmetry $S$ as the
ratio between the areas of violet and red peaks of the emission line.
Let us highlight the principal features of the resulting curves\footnote{Due
to the wavelength response of the detectors used during the first observations,
the signal-to-noise ratio (S/N) of H$\beta$ is significantly higher than
of the other lines. Therefore, this discussion is mainly based on the
H$\beta$ line. However, these results are completely consistent with
the behaviour of H$\gamma$ and H$\delta$.} (Fig.~\ref{swave}).
Note first that their shape obviously deviates from of a sine wave,
although for an accretion disk with a single bright spot it should be
nearly sinusoidal (Borisov and Neustroev \cite{bor:neus1}).
In addition to this, the amplitude of this intensity variation of
the blue peak exceeds by far that for the red peak.
However, it can be seen that the degree of asymmetry with $S > 1$ is observed only during
one-half of the orbital period, whereas $S \leq 1$ is found at other times.
The possible reasons for such variations of the degree of asymmetry
can be an anisotropic radiation of the bright spot \textit{and/or} an eclipse of the
bright spot by the outer edge of the accretion disk.
These assumptions will be discussed in the course of further analysis.

A closer inspection of the line profiles shows that
all the emission lines exhibit a considerable asymmetry of their wings which
varies with time. This effect is observed in both data sets but most
notably in the first one. The visual analysis has shown that from time to time
a hump appears on the blue and red wings (Fig.~\ref{wing_as}).
Its shift concerning the center of the line does not exceed
$\sim$1200--1400 km~s$^{-1}$.

To describe the wing asymmetry WA quantitatively, we define the following quantity:

\begin{equation}
WA = \left\{ {\begin{array}{*{20}c}
   {VR - 1} & {\begin{array}{*{20}c}
   {\hspace{1.5cm} VR > 1} \\
\end{array}} \\
   {1 - VR^{ - 1} } & {\begin{array}{*{20}c}
   {\hspace{1.5cm} VR < 1} \\
\end{array}} \\
\end{array}} \right.
\end{equation}

\noindent where
\[
VR = {{\sum\nolimits_{\lambda_0 - d\lambda }^{\lambda_0 - (d\lambda + \Delta\lambda )} {I_\lambda}}
     \mathord{\left / \,{\vphantom {{\sum\nolimits_{\lambda_0 - d\lambda}^{\lambda_0 - (d\lambda +
     \Delta\lambda)} {I_\lambda }}{\sum\nolimits_{\lambda_0 + \lambda_b }^{\lambda_0 +
     (d\lambda + \Delta\lambda )} {I_\lambda } }}} \right.
     \kern-\nulldelimiterspace} {\sum\nolimits_{\lambda_0 + d\lambda }^{\lambda_0 +
     (d\lambda + \Delta\lambda )} {I_\lambda } }}
\]

\noindent
Here ${\lambda_0 }$ is the wavelength of the center of the line, and $d{\lambda}$
defines the beginning of the wavelength interval ${\Delta\lambda}$
in which the fluxes are summed up.

\begin{figure*}
\centering
\resizebox{15cm}{!}{\includegraphics{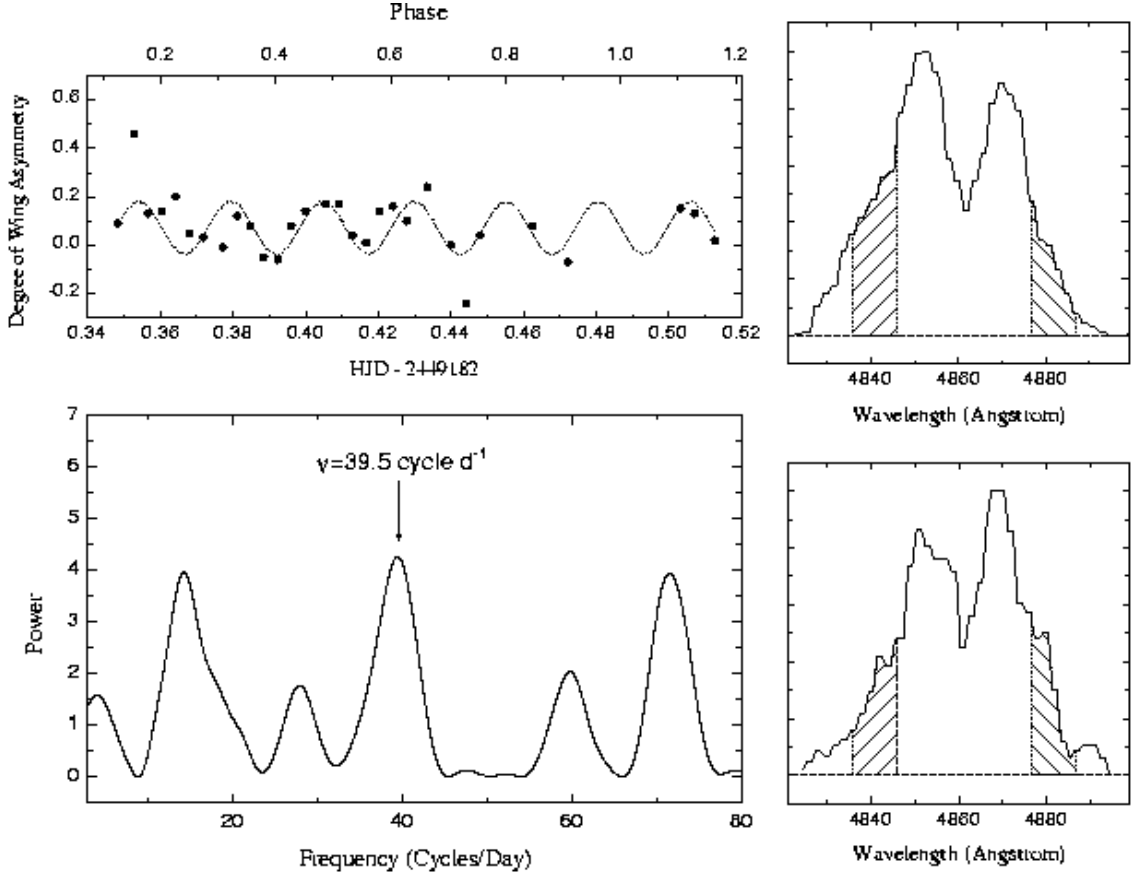}}
\caption{
In the upper left panel is shown the degree of the wing asymmetry of the emission
line H$\beta$ from the first dataset plotted against the time of observations.
These data have been fitted by the simple sinusoid with the period 0\fh61
(dotted line).
In the lower left panel the periodogram of the above data is shown.
The periodogram shows three strong peaks at 0\fh61, 0\fh33 and 1\fh67,
the first one is the little bit stronger of them.
In the right-hand column are shown some profiles of the H$\beta $ line
in which the wavelength range used for the determination of the degree
of the wing asymmetry has been marked.}
\label{wing_power}
\end{figure*}

To see how the wing asymmetry of the emission lines depends on time, we
calculated the degree of the wing asymmetry of all H$\beta$ profiles and
plotted their values against the time of observations
(Fig.~\ref{wing_power}, upper left panel).
In this case we have used following parameters:
$d{\lambda}$ = 15~\AA\ and ${\Delta\lambda}$ = 10~\AA.
It can be seen that the wing asymmetry shows quasi-periodic modulations with
a period much shorter than the orbital one. This indicates the presence
of an emission source rotating asynchronously with the binary system.

We looked for periods in this data sample using the Scargle (\cite{Scargle})
method and tested our results using the AoV method (Schwarzenberg-Czerny \cite{Sch-Cz})
and the PDM algorithm (Phase Dispersion Minimization,
Stellingwerf \cite{Stellingwerf}).

The calculated periodogram shows three strong peaks at 0\fh61, 0\fh33 and 1\fh67,
the first one being little bit stronger of them
(Fig.~\ref{wing_power}, lower left panel).
This result was confirmed applying the AoV and PDM methods.
As an example we have fitted the wing asymmetry's data from
Fig.~\ref{wing_power}
(upper left panel) by a simple sinusoid with the period of 0\fh61 and
have plotted this curve by a dotted line in the same figure.

A detailed discussion of the line-wings` variations is given in Sect.~\ref{Discussion}.

\section{Investigation of the structure of the accretion disk}
\label{structure}

The orbital variations of the emission line profiles indicate a
non-uniform structure of the accretion disk. For its study we have used the
Doppler tomography and the Phase Modelling Technique.

\subsection{The Doppler tomography}
\label{Doppler}

Doppler tomography is an indirect imaging technique which can be used to
determine the velocity-space distribution of the line's emission in close binary
systems. Full details of the method are given by Marsh \& Horne
(\cite{marsh:horne}).
Doppler maps of the major emission lines were computed using the filtered
backprojection algorithm
described by Robinson et al. (\cite{robinson}).
Since we cannot assume any relation between position and velocity during
eclipse we constrained our data sets by removing eclipse spectra covering
the phase ranges $\varphi = 0.90 - 0.10$.

\begin{figure*}
\centering
\resizebox{15cm}{!}{
\includegraphics{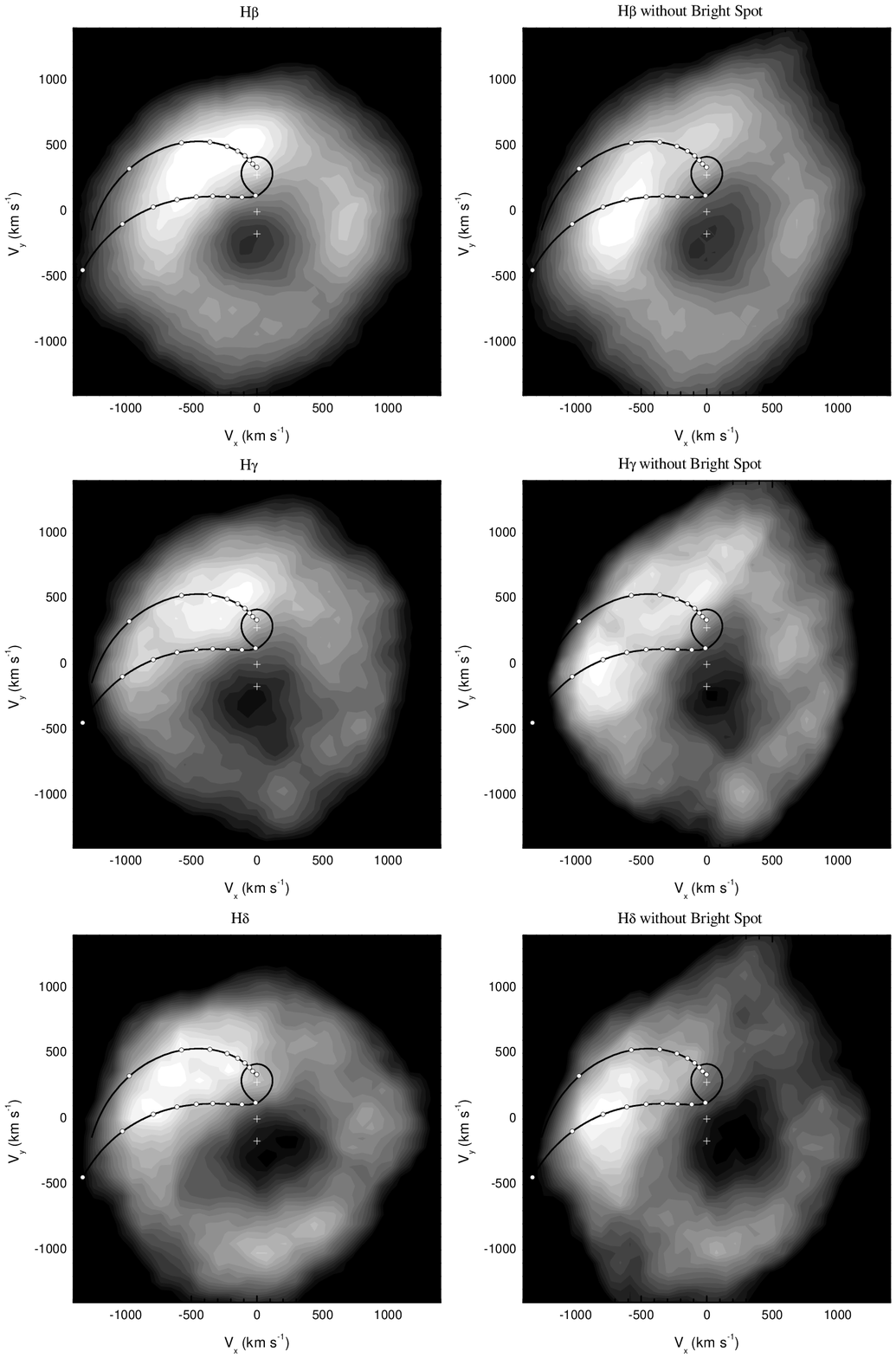}}
\vspace*{0.5cm}
\caption{
Doppler maps of H${\beta }$, H${\gamma}$ and H${\delta }$ emission obtained
from the first dataset. Left column: Doppler maps based on all spectra.
Right column: the maps based on those spectra which were obtained at minimum
spot brightness (for details see text).
}
\label{my_dop}
\end{figure*}

\begin{figure*}
\centering
\resizebox{15cm}{!}{
\includegraphics{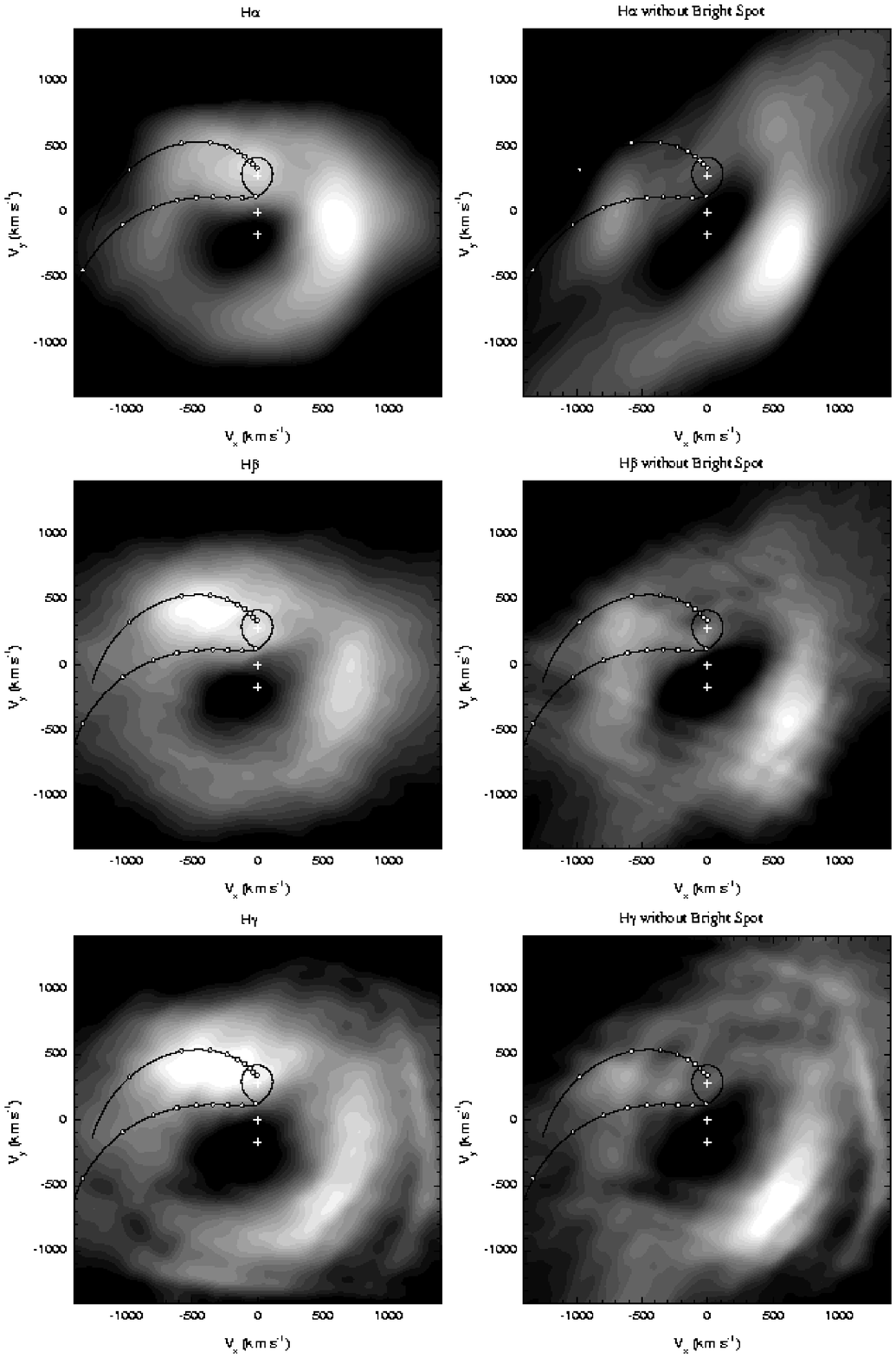}}
\vspace*{0.5cm}
\caption{
Doppler maps of H${\alpha}$, H${\beta }$ and H${\gamma}$ emission obtained from the
second dataset. Left column: Doppler maps based on all spectra.
Right column: the maps based on those spectra which were obtained
at minimum spot brightness (for details see text).
}
\label{bob_dop}
\end{figure*}

The constructed Doppler maps of the H$\beta$, H$\gamma$ and H$\delta$
emission from the first dataset are shown in Fig.~\ref{my_dop} (left column).
Though the tomograms based on the spectra from the second dataset,
were already published, we find it useful to present here again the Doppler maps of H$\alpha$,
H$\beta$ and H$\gamma $ (Fig.~\ref{bob_dop}, left column).
The figures also show the positions of the white dwarf, the center of mass of the
binary and the secondary star, and also the trajectories of free particles released
from the inner Lagrangian point. Additionally, the velocity of the disk along the
path of the gas stream is plotted in each Doppler map.

The tomograms show at least the two bright
emitting regions superposed on the typical ring-shaped emission of the
accretion disk. The bright emission region with coordinates
$V_{x} \approx$ 0 $-$ -800 km s$^{-1}$ and
$V_{y} \approx$ 0 $-$ 700 km s$^{-1}$ (first dataset)
and V$_{x} \approx$ -200 $-$ -800 km s$^{-1}$ and
$V_{y} \approx$ 200 $-$ 600 km s$^{-1}$ (second dataset)
can be unequivocally contributed
to emission from the bright spot on the outer edge of the accretion disk.
The second spot with coordinates
$V_{x} \approx$ 700 $-$ 800 km s$^{-1}$ and $V_{y} \approx$ 50
km s$^{-1}$
(first dataset, most notable in H${\beta }$ and H${\delta }$) and
$V_{x} \approx$ 200 $-$ 800 km s$^{-1}$ and $V_{y} \approx$
-900 $-$ 300 km s$^{-1}$ (second dataset)
locate far from the region of interaction between the stream
and the disk particles. The nature of this spot is analysed in the following section.

\subsection{The modelling}
\label{modelling}

Although the Doppler tomography is a very robust technique which can analyse the
structure of the accretion disk, it still has some imperfections.
We shall note only one of them. This technique makes no allowance for changes
in the intensity of features over an orbital cycle. Components which do vary will be
handled incorrectly, and the obtained tomograms will be averaged on the period.
In our recent paper (Borisov \& Neustroev \cite{bor:neus1}) we have presented
another technique for the investigation of the structure of the accretion disk,
which is based on the modelling of the profiles of the emission lines formed in a
non-uniform accretion disk. The analysis has shown that the modelling of the spectra
obtained in different
phases of the orbital period, allows to estimate the principal parameters of the spot,
though its spatial resolution is worse.
However, there is also an important advantage --- the method
allows to investigate the modification of the spot's brightness with orbital phase.
This cannot be done by Doppler tomography.

For an accurate calculation of the line profiles formed in the
accretion disk, it is necessary to know the velocity field of the radiating gas,
its temperature and density, and, first of all, to calculate the radiative
transfer
equations in the lines and the balance equations. Unfortunately, this complicated
problem has not been solved until now and it is still not possible to reach an
acceptable consistency between calculations and observations. Nevertheless,
even the simplified models allow one to derive some important
parameters of the accretion disk.

\begin{figure}
\resizebox{\hsize}{!}{\includegraphics{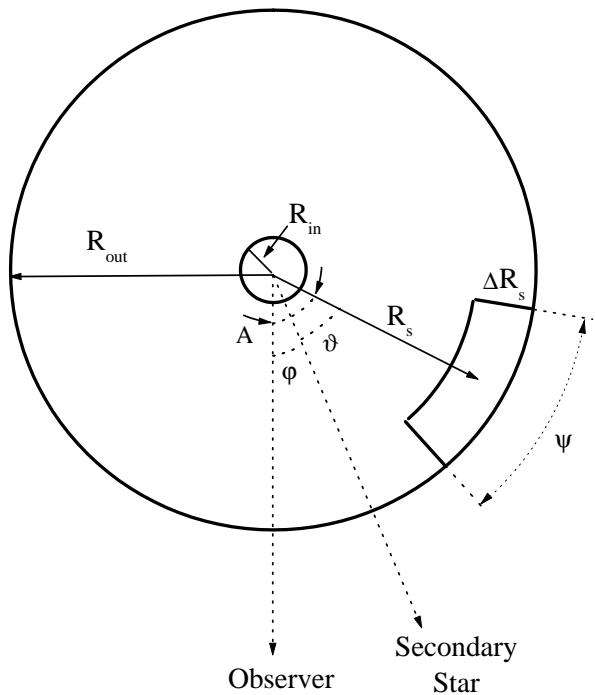}}
\caption{Accretion disk geometry for the line profile
model described in the text.}
\label{model}
\end{figure}

In our calculations we have
applied a double-component model which include the flat Keplerian geometrically
thin accretion disk and the bright spot whose position is constant
with respect to components of the binary system (Fig.~\ref{model}).
We began the modelling of the line profiles
with calculation of a symmetrical double-peaked profile formed in the uniform
axisymmetrical disk, then we added the distorting component formed in the bright
spot.
To calculate the line profiles we have used the method of Horne \& Marsh
(\cite{Horne:Marsh}), taking into account the Keplerian velocity gradient across
the finite thickness of the disk.

Free parameters of our model are:
\begin{enumerate}
\item $R=R_{in}$/R$_{out}$ is the ratio of the inner and outer radii of the disk;
\item $V$ is the velocity of the outer rim of the accretion disk;
\item $\alpha$ is an emissivity parameter (the line surface brightness of the disk is
assumed
to scale as R$^{-\alpha}$);
\item $\vartheta$ is an angle between the direction from the primary to the
secondary and the center of the bright spot respectively (a phase angle);
\item $\Psi$ is a spot azimuthal extent;
\item $R_S$ is a radial position of the spot's center in fractions of the outer
radius ($R_{out}$=1);
\item $\Delta R_S$ is a radial extent;
\item $L$ is a relative dimensionless luminosity of the spot,
 which is given by

\vspace*{0.5cm}
\noindent$
\begin{array}[t]{l}
L=\int\limits_{R_{S}-\Delta
R_{S}/2}^{R_{S}+\Delta R_{S}/2}S\cdot B\cdot f(r)\cdot dr= \\ \\
=\frac{\pi}{180}\frac{\Psi \Delta R_{S}B}{2-\alpha}\left[\left(R_{S}+
\frac{\Delta R_{S}}{2}\right)^{2-\alpha}-\left(R_{S}-\frac{\Delta R_{S}}{2}
\right)^{2-\alpha}\right]
\end{array} $

\mathstrut
\noindent where $S$ is area of the spot, and $B$ is the spot's contrast
(the spot-to-disk brightness ratio at the same distance from the white dwarf).
\end{enumerate}

The results of testing have shown (Borisov and Neustroev \cite{bor:neus1})
that for a reliable estimation of the parameters it is important to keep the certain sequence
of their determination. This is connected with the strong azimuth dependence of the emission
line profile with variation of various spot parameters.
When modelling the emission lines of IP Peg, we determined the
parameters in the following order:

\begin{itemize}
\item[\textbullet] The orbital-phase modulation of the degree of asymmetry of the
profile is used to find the phase angle of the spot $\vartheta$;
\item[\textbullet] Spectroscopic data are then sorted according to the "maximally effective"
phases for various parameters of the spot: the azimuthal extent of the spot $\Psi$ can be best
determined at azimuths -$30^{\circ} - 30^{\circ}$ and $150^{\circ} - 210^{\circ}$,
and the radial position of the spot $R_S$ can be best determined at azimuths
$50^{\circ } - 130^{\circ}$ and -$50^{\circ }-$ -$130^{\circ }$;
\item[\textbullet] The azimuthal extent of the spot is determined, fixed, and then used in the
estimation of the radial position of the spot in the accretion disk;
\item[\textbullet] The remaining profiles are modelled with values of the already determined
geometrical spot parameters fixed,
in order to derive the relative luminosity of the spot at various orbital phases.
\end{itemize}

\begin{table*}[t]
\caption[] {Average parameters of the accretion disk and the bright spot obtained by
modelling separate spectra}
\begin{tabular}{clcccccccc}
\hline
\hline
\noalign{\medskip}
Emission & Dataset & $\alpha$ & $R$ & $V$ & $\vartheta$ & $\Psi $ & $\Delta R_{s}^{\ast}$ & $R_{s}$ & Contrast \\
line & & & & & & & & & \\
\hline\noalign{\smallskip}

H${\beta }$ & First & 1.67 $\pm$ 0.49 & 0.08 $\pm$ 0.04 & 570 $\pm$ 100 & 30 & 50 $\pm$ 15 & 0.1 & 0.95 $\pm$ 0.05 & 2.1 $\pm$ 1.5 \\
H${\beta }$ & Second & 1.61 $\pm$ 0.40 & 0.08 $\pm$ 0.03 & 590 $\pm$ 106 & 29 & 49 $\pm$ 18 & 0.1 & 0.96 $\pm$ 0.05 & 2.0 $\pm$ 1.1 \\
H${\gamma }$ & First & 2.15 $\pm$ 0.56 & 0.08 $\pm$ 0.03 & 549 $\pm$ 130 & 30 & 75 $\pm$ \ 6 & 0.1 & 0.90 $\pm$ 0.10 & 3.1 $\pm$ 2.6 \\
H${\gamma }$ & Second & 1.69 $\pm$ 0.47 & 0.08 $\pm$ 0.04 & 616 $\pm$ 102 & 26 & 42 $\pm$ \ 5 & 0.1 & 0.89 $\pm$ 0.08 & 4.9 $\pm$ 4.9 \\
H${\delta }$ & First & 1.68 $\pm$ 0.78 & 0.12 $\pm$ 0.06 & 669 $\pm$ 155 & 30 & 52 $\pm$ 16 & 0.1 & 0.88 $\pm$ 0.08 & 3.6 $\pm$ 3.2 \\
\noalign{\smallskip}
\hline
\noalign{\bigskip}
\end{tabular}
\\
$\ast$ Adopted by default.
\label{Table1}
\end{table*}

Since the shape of the profile is virtually independent of variations in the
radial extent of the spot $\Delta R_S$ (Borisov \& Neustroev \cite{bor:neus1}),
a default value for this parameter can be used in
the profile computations (e.g., a typical radial extent of the bright spot).
Photometric observations of the cataclysmic variables indicate that the radial extent
of the spots vary from ~0.02 to ~0.15 (Rozyczka~\cite{Rozyczka}).
We adopted the value $\Delta R_S$ = 0.1.

The necessary condition for the accurate determination of the
spot parameters is the knowledge of its phase angle $\vartheta$, which possibly
can be found from the analysis of the phase variations of the degree of asymmetry
of the emission line (S-wave graph).
We must determine $\varphi_{0}$. This phase corresponds to the moment when the
radial velocity of the S--wave component is zero. Then the phase angle $\vartheta$
of the spot will be $2 \pi (1 - \varphi _{0})$.
Examples of the application of this technique to real data are given in Neustroev
(\cite{neustroev}) and Borisov \& Neustroev (\cite{bor:neus2}).

For the study of the accretion disk structure of IP Peg it was possible to
apply this technique to all major emission lines (H$\alpha$, H$\beta$, H$\gamma$ and
H$\delta$), but because of signal-to-noise limitations the discussion, following
below, is mostly based on the H$\beta$ line. We have started our analysis of the
spectroscopic data by determination of the phase angle $\vartheta$ of the spot,
using the S-wave graph (Fig.~\ref{swave}). The value of $\vartheta$ has been found
to be 30$^{\circ }$ (H$\beta$, H$\gamma$ and H$\delta$ lines from first dataset),
29$^{\circ }$ (the H$\beta$ line from second dataset) and 26$^{\circ}$ (the H$\gamma$
line from second dataset) respectively.

As a result of subsequent modelling we have established all principal
parameters of the accretion disk and the bright spot.
Their average values are listed in Table~\ref{Table1}.
It is necessary to note that the standard deviations of all averaged parameters
of the disk are considerably higher than expected on the basis of testing
(Borisov \& Neustroev \cite{bor:neus1}). It may be due to possible orbital
variations of the observed parameters of the accretion disk.

\begin{figure}
\resizebox{\hsize}{!}{\includegraphics{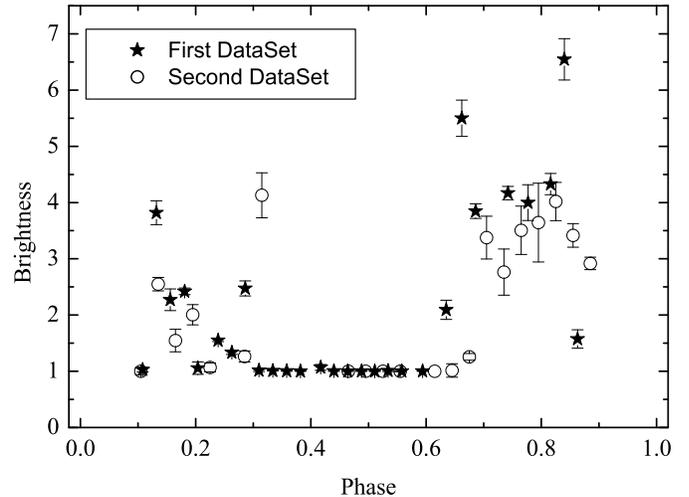}}
\caption{
The dependence of the spot brightness on orbital phase, obtained from
modelling of the H$\beta$ emission line. Obviously the brightness
considerably oscillates, and during a significant part of the period the spot is not
visible.
}
\label{brightness}
\end{figure}

\section{Discussion}
\label{Discussion}

As already noted in Sect.~\ref{Doppler}, our Doppler maps show
two bright emitting regions (Figs.~\ref{my_dop} and \ref{bob_dop}, left column).
Now we also point out that the part of the bright region which we have
interpreted as bright spot, is located near the intersection of a ring-shaped structure
and the upper arc of the tomograms. It indicates the nearly Keplerian velocities
of the matter in the area of the bright spot.

The modelling of the Balmer line profiles has allowed us to find the dependence
of the brightness of this spot on the orbital phase (Fig.~\ref{brightness}).
The derived brightness curves of the spot for both observing sets are very similar.
One can see that the brightness considerably oscillates, and during a significant part
of the period ($\varphi \sim$ 0.2 to 0.6) the spot is not visible.
This happens when the spot is on the distant
half of the accretion disk. On the contrary, the spot becomes brightest at the
moment of inferior conjunction.
It is important to note that even near to the moment of inferior conjunction
the brightness of the spot varies. Probably, this is connected to an
anisotropic radiation of the bright spot. In consequence of it we get an
asymmetric change in radiation during one orbital revolution of the system.
However, the anisotropy alone cannot explain the missing emission of
the bright spot during the orbital phases when the spot has turned away
from the observer's point of view. (In this case it is more correct to name
"a bright spot" as "an invisible hot spot".) We think this fact can
additionally be explained by a self-eclipse of the bright spot owing to
the large inclination of the orbital plane of IP Peg and of its accretion disk.
This scenario causes an eclipse of the bright spot by an outer edge of
the accretion disk followed by a drop of observable spot brightness.
Thus we suggest at least two mechanisms for explanation of the observable
variations of the spot brightness: an anisotropic radiation of the bright
spot and an eclipse of the bright spot by the outer edge of the accretion disk.
We have no possibility to discuss in this paper, which of these mechanisms
is more important. It will be the subject of the separate paper.

\begin{figure}
\resizebox{\hsize}{!}{\includegraphics{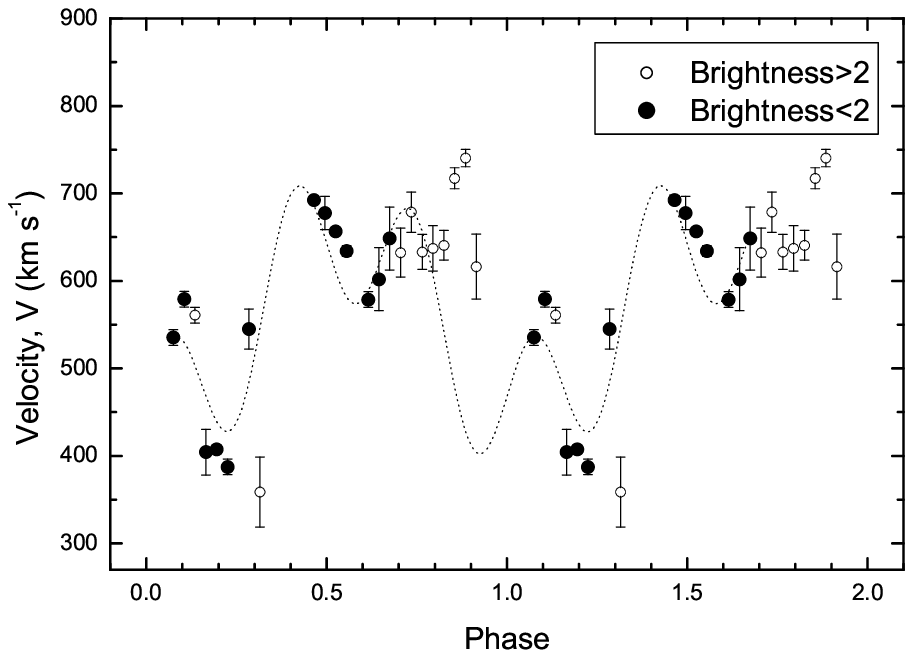}}
\resizebox{\hsize}{!}{\includegraphics{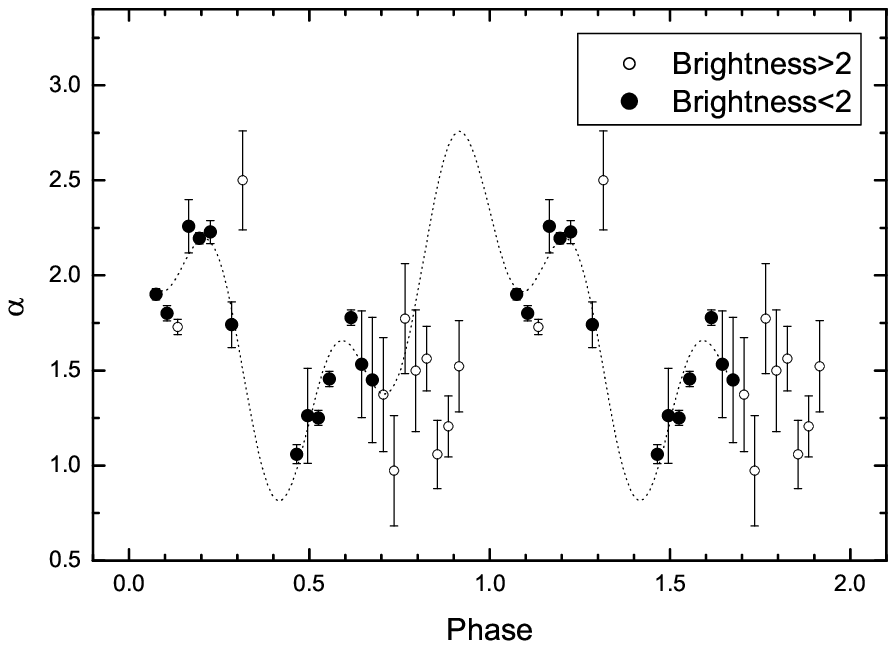}}
\caption{
The dependence of the velocity of the outer edge of the accretion disk
(top panel) and of the emissivity parameter $\alpha$ (bottom panel) on
orbital phase, obtained on basis of modelling H$\beta$ emission line from
the second dataset. Filled and open circles represent those values of parameters,
which were found on spectra obtained in phases of minimum and maximum brightness
of the spot respectively.
}
\label{velocity}
\end{figure}

The second area of increased luminosity is located too far from the region of
interaction between the stream and the disk particles. None of the theories do predict
here the presence of a bright spot, which is connected with such an interaction.
This area was interpreted recently by Wolf et al. (\cite{Wolf98}) as beginning
of the formation of a spiral arm in the outer disk in context of the tidal force
from the secondary.

The majority of the simulations predicts the presence of
the spiral structure with two symmetrically located spiral shocks in the
accretion disk.
Exactly such a two-armed structure was detected by Steeghs et al. (\cite{Steeghs97})
in the accretion disk of IP Peg during outburst.
However, the second spiral arm in our tomograms is not visible.
If it does exist, it is probably hidden in the intensive
emission of the bright spot.
If the contribution of the bright spot from the initial
profiles of the emission lines (or from the resulting Doppler maps),
could be removed, it may be possible to detect the second spiral arm.

For this we have constructed the new Doppler maps using only half of the spectra
obtained in phases of minimum brightness of the spot ($\varphi$ = 0.15
to 0.65) (Figs.~\ref{my_dop} and \ref{bob_dop}, right column).
Although the used profiles of the emission lines
practically do not contain information about the bright spot, on the left
side of the Doppler maps we can see an area of increased luminosity! However
its location has changed. It was displaced downwards and has taken practically
a symmetrical position concerning the second area of the increased luminosity.
However, in contradiction to Steeghs et al (\cite{Steeghs97}),
we cannot confidently assert that the form of both bright areas is spiral,
first of all because of their low brightness. Indeed, the emissivity contrast
between the proposed spirals and other parts of the disk is less than 1.3
for all emission lines. At the same time we consider
that even the really spiral features in the tomograms can be caused by a different effect
than the spiral shocks (see, for example, Smak~\cite{Smak01}; Ogilvie~\cite{Ogilvie}).
Additional evidences for this are necessary.

\begin{figure*}
\centering
{
\includegraphics[width=8cm]{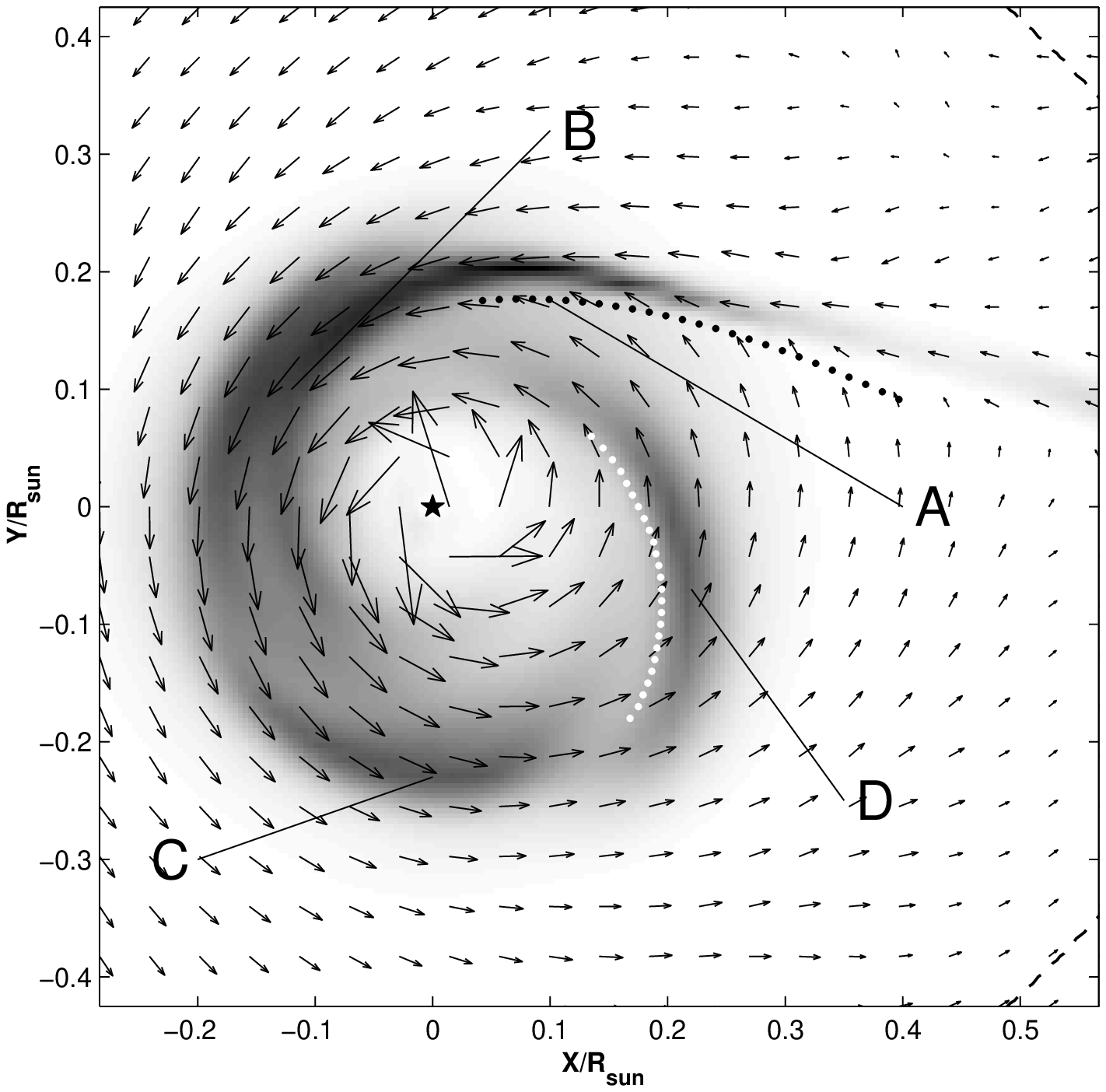}
\includegraphics[width=8.13cm]{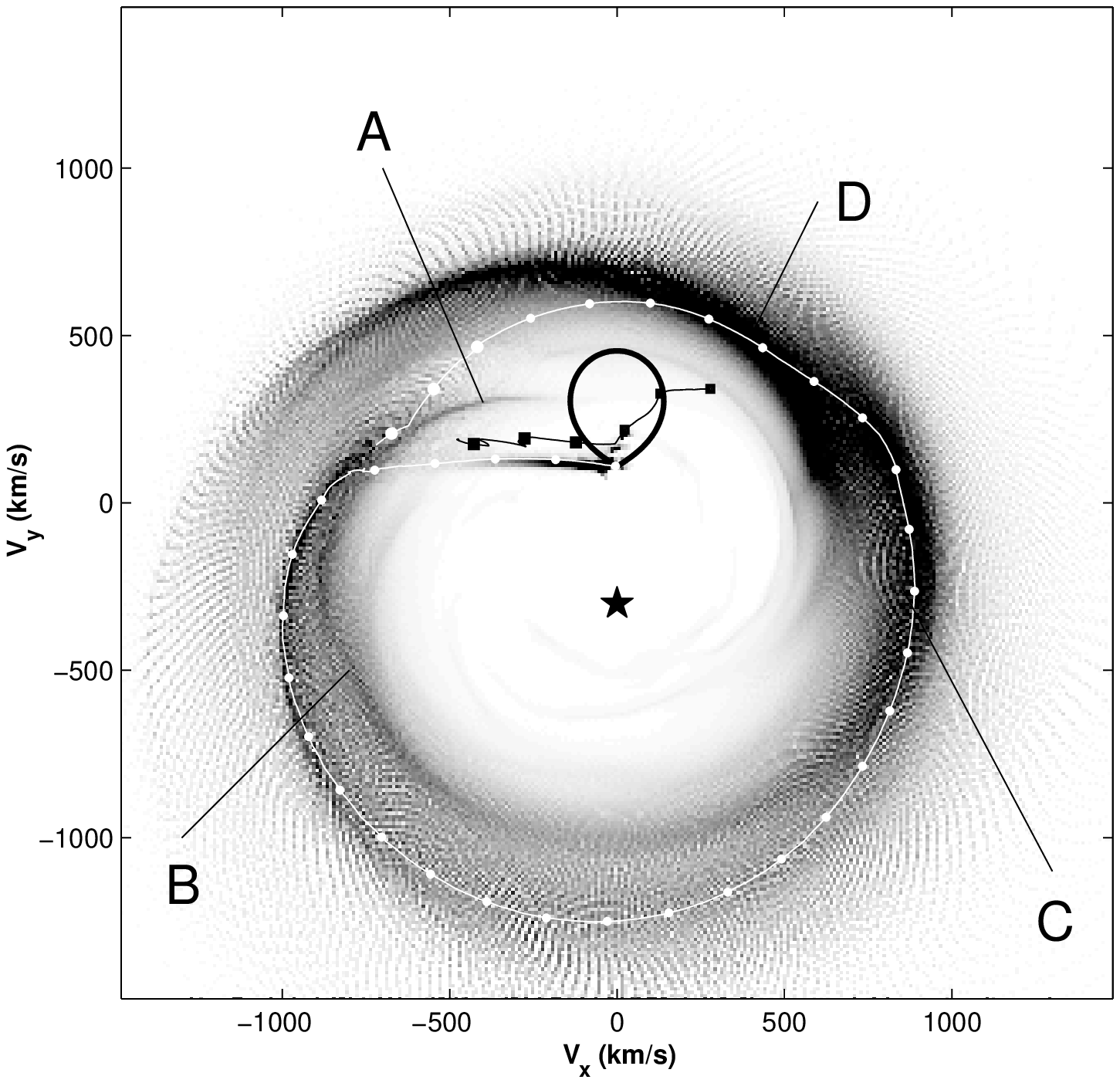}
}
\caption{\textit{Left:} The distribution of $\rho^{2}T^{1/2}$ over the equatorial
plane in 3D calculations of Kuznetsov et al. (\cite{Kuznetsov})
($\rho$ is density, $T$ is temperature). Arrows are the velocity
vectors in observer's frame. The asterisk is the white dwarf. The white
dotted line is the tidally induced spiral shock. A black dotted line is the
shock wave along the edge of the stream (`hot line'). The main emission
regions are marked by \textbf{A, B, C, D}.
\textit{Right:} Synthetic Doppler map for $I\sim \rho^{2}T^{1/2}$. The secondary
Roche lobe (a bold black line) and the white dwarf (an asterisk) are also
shown. The white line with circles and black line with squares show gas dynamical
trajectories in the velocity coordinates. The main emission regions are marked
by \textbf{A, B, C, D} as above. Both figures are
reproduced under the kind permission by O.A. Kuznetsov and D.V. Bisikalo.}
\label{bisikalo}
\end{figure*}

Additional evidences for a spiral structure of the accretion disk of IP Peg arise from
the dependence of the velocity of the outer edge of the disk on orbital phase.
Earlier, studying the structure of the accretion disk of U~Gem, we have detected a
sinusoidal variation of the parameters of the disk (and especially the velocities V of
its outer edge) of the form $\sin 2\varphi $. We interpreted this by the presence of a spiral
structure (Neustroev \& Borisov \cite{Neus:Bor}) in the disk of
U Gem.
We have tried to detect similar effects in IP Peg. In Fig.~\ref{velocity} (top panel)
the dependence of the velocity of the accretion disk on
orbital phase is shown.
One can see, that the amplitude of variation of V in this case is even higher
than for U Gem. However, the most significant variations of velocity occur
at the moment of maximum brightness of the spot. In this case we cannot
unequivocally eliminate the probable influence of the bright spot on the double peak
separation of the emission lines.
Therefore in Fig.~\ref{velocity} we have selected those values of
velocity V, which were found from spectra obtained in phases of minimum
brightness of the spot. Such a spot distorts the line profile only marginally,
therefore the velocity V is determined confidently.

Although the dependence of the velocity V on orbital phase, obtained from
the first dataset, does not allow us confidently to reveal
any regularity in modification of the velocity V, the second dataset confidently
indicates the deviation from the Keplerian velocity field in the accretion disk
of IP Peg (Fig.~\ref{velocity}). It is important to note that other parameters of the disk
also vary as $\sin 3\varphi$ simultaneously with V.
Moreover, the explicit anti-correlation between a change of V and $\alpha $ is
observed, the correlation coefficient being more than 0.96 with a
confidence probability of 99\%.
Variations of the peak separation of the hydrogen lines of the
form $\sin 3\varphi $ is the signature of an m=3 mode in the disk.
This mode can be excited, also as in U~Gem, by the tidal forcing
(whose main component is the m=2 mode) and the detected variations
can also be explained by the presence of spiral shocks in the accretion disk.

The presence of the compact zone of energy release in the area of interaction of
the stream and the accretion disk and the spiral shocks in the disk of IP Peg
in quiescence does not contradict the most accurate modern 3D gasdynamical models
(Bisikalo et al.~\cite{Bisikalo98a}, \cite{Bisikalo98b};
Makita et al.~\cite{Japan00}). Let us consider the results of these investigations
in more detail in application to IP Peg.
Kuznetsov et al. (\cite{Kuznetsov}) presented synthetic Doppler maps of
gaseous flows in this system based on the results of their 3D gasdynamical
simulations. They concluded that there are four elements of the flow
structure which contribute to the total system luminosity: the `hot line' \footnote{
In calculations of Bisikalo et al. (\cite{Bisikalo98a}, \cite{Bisikalo98b}) and
Makita et al. (\cite{Japan00}), the gas stream from $L_{1}$ did not cause
the shock perturbation of the disk boundary. It meant the absence of the `hot spot'.
At the same time, the gas of the `circumbinary envelope' interacts with the stream and
causes the formation the extended shock wave, located on the stream edge (`hot line').}
(region \textbf{A} in Fig.~\ref{bisikalo}), the most luminous part of the stream where
the density is still large enough and the temperature already increases due to
dissipation (region \textbf{B}), the dense region near the
apastron of the disk (region \textbf{C}), and the dense post-shock region attached to the
spiral shock (region \textbf{D}). The income of each element obviously can
vary depending on the peculiarities of the considered binary system.\footnote{
Bisikalo et al. (\cite{Bisikalo98a}, \cite{Bisikalo98b}) also claimed that we see only
the one-armed
spiral shock. In the place where the second arm should be (region \textbf{B}) the stream
from $L_{1}$ dominates and presumably prevents the formation of second arm of tidally
induced spiral shock. The last claim is disputed by Makita et al. (\cite{Japan00}) and
this problem remains unclosed. From the observation side these two versions practically
are indiscernible.}
The comparison of the observational tomograms from Figs.~\ref{my_dop} and \ref{bob_dop}
and with synthetic ones from Fig.~\ref{bisikalo} reveals that the dominating
elements in the accretion disk of IP Peg are the `hot line', the dense zone near
the disk's apastron (region \textbf{C}) and the post-shock zone attached to the
arm of a spiral
shock (region \textbf{B}). Signatures of a spiral shock in the region \textbf{D}
are not detected.

Thus we believe that our observations as a whole confirm the spiral structure of
the quiescent accretion disk of IP Peg.
At the same time we point out that the determined structure of the accretion
disk of IP Peg satisfies modern 3D simulations, which predict a
considerably more complicated structure of the accretion disk than found by earlier
calculations.
Note that Marsh \& Horne (\cite{marsh:horne90}) and Harlaftis et al.
(\cite{Harlaftis94}) have also obtained the tomograms of IP Peg which are very
similar to ours, indicating that the detected spiral structure of the IP Peg's
accretion disk is long-lived structure.

In conclusion we would like to discuss possible reasons for the detected variability
of the wings of the emission lines. The results of 3D numerical simulations of
Bisikalo et al. (\cite{Bisikalo01a}, \cite{Bisikalo01b}) have shown that if
spiral shocks are present in the accretion disk then \textit{any} disturbance
of the disk would result in the appearance of a blob, the later moving through the
disk with variable velocity but with constant period of revolution.
Generally the period of the blob revolution depends on the value of viscosity,
and for typical accretion disks with $\alpha \sim 0.01 - 0.1$ the period
should be in the range of $0.10 - 0.20 P_{orb}$. This dense formation lives long
enough and retains its main characteristics for a time of the order of tens orbital
periods. Furthermore, every new disturbance of the disk structure will transform
into a blob.\footnote{It is fair to say that Bath (\cite{Bath}) was the first who
suggested the presence of the blobs in the accretion disks,
though the Bath's blobs are quite different than the one described here.}

If such a blob exists, it should distort the profiles of the spectral lines
with the period of the blob's revolution. From Figs. 3 and 4 in Bisikalo et al.
(\cite{Bisikalo01b}) it can be seen, that the distance from the blob to the
white dwarf is about 0.1 of the binary separation. Hence maximum radial
velocity of the line component which forms in the blob should be close to
$\pm$1200 km s$^{-1}$. Consequently, best areas of the spectral line for detecting
of the blob should be the line wings. Thus, the detected variability of
the wings of the emission lines with the period of 0\fh61 ($\sim$0.15$P_{orb}$) can be
really explained by a blob. Furthermore, there is additional evidence for
the presence of
spiral shocks in the accretion disk of IP Peg, as the existence of the blob
is maintained by the spiral shocks.

\section{Conclusions}
\label{Conclusions}

In this paper we have presented the results of spectral investigations of the
cataclysmic variable IP Pegasi in quiescence. The analysis of optical spectra
by means of the Doppler tomography and the Phase Modelling Technique has allowed us
to make the following conclusions:
\begin{enumerate}
\item
The quiescent accretion disk of IP Peg has a complicated structure.
Together with the bright spot, originated in the region of interaction between
the stream and the disk particles, there are also explicit indications of spiral shock waves.
The Doppler maps, the variations of the peak separation of the emission lines and
the variability of the asymmetry of the line wings are indicative.
\item
The brightness of the bright spot considerably oscillates, and during a significant part of
the period the spot is not visible. This happens when the spot is on the distant
half of the accretion disk. On the contrary, the spot becomes brightest at the
moment of an inferior conjunction.
Probably, it is connected to an anisotropic radiation of the bright
spot and an eclipse of the bright spot by the outer edge of the accretion disk.

\item
The material located in the area of the bright spot on the accretion disk, is
moving with nearly Keplerian velocities.
\item
All the emission lines show a rather considerable asymmetry of their wings which
varies with time. The wing asymmetry
shows quasi-periodic modulations with a period much shorter than the orbital one.
This indicates the presence of an emission source rotating asynchronously
with the binary system.
\end{enumerate}

\begin{acknowledgements}

VVN would like to thank the firm VEM (Izhevsk, Russia) and Konstantin Ishmuratov
personally for the financial support in the preparation of this paper.
Thanks also to Alexander Khlebov for the computer and technical support.
We thank Dmitry Bisikalo, Takuya Matsuda and Makoto Makita for discussions on
this topic. We acknowledge the referee Dr. P. Godon for detailed reading
of the manuscript and useful suggestions concerning the final version.

\end{acknowledgements}

\end{document}